\documentclass{vgtc}

\ifpdf
  \pdfoutput=1\relax
  \usepackage{graphicx}
  \DeclareGraphicsExtensions{.pdf,.png,.jpg,.jpeg}
\else
  \usepackage{graphicx}
  \DeclareGraphicsExtensions{.eps}
\fi%

\graphicspath{{figures/}{pictures/}{images/}{./}}

\usepackage{microtype}
\PassOptionsToPackage{warn}{textcomp}
\usepackage{textcomp}
\usepackage{mathptmx}
\usepackage{times}

\usepackage{cite}

\usepackage{enumitem}

\onlineid{0}

\vgtccategory{Research}

\vgtcinsertpkg

\title{Toward a Bias-Aware Future for Mixed-Initiative Visual Analytics}

\author{Adam Coscia\thanks{e-mail: acoscia6@gatech.edu}
\and Duen Horng Chau\thanks{e-mail: polo@gatech.edu}
\and Alex Endert\thanks{e-mail: endert@gatech.edu}}
\affiliation{\scriptsize Georgia Tech}

\abstract{
Mixed-initiative visual analytics systems incorporate well-established design principles that improve users' abilities to solve problems.
As these systems consider whether to take initiative towards achieving user goals, many current systems address the potential for cognitive bias in human initiatives statically, relying on fixed initiatives they can take instead of identifying, communicating and addressing the bias as it occurs.
We argue that mixed-initiative design principles can and should incorporate cognitive bias mitigation strategies directly through development of mitigation techniques embedded in the system to address cognitive biases in situ.
We identify domain experts in machine learning adopting visual analytics techniques and systems that incorporate existing mixed-initiative principles and examine their potential to support bias mitigation strategies.
This examination considers the unique perspective these experts bring to visual analytics and is situated in existing user-centered systems that make exemplary use of design principles informed by cognitive theory.
We then suggest informed opportunities for domain experts to take initiative toward addressing cognitive biases in light of their existing contributions to the field.
Finally, we contribute open questions and research directions for designers seeking to adopt visual analytics techniques that incorporate bias-aware initiatives in future systems.
}

\begin{document}

\firstsection{Introduction}

\maketitle

Mixed-initiative systems provide a collaborative approach to performing tasks that balance human and machine effort. 
This design metaphor exists in visual analytics (VA) systems in the form of machine-learning (ML) models and algorithms that seek to combine human-in-the-loop goals with techniques for automation while leveraging the power of visualization for creating insights from information.
Horvitz \cite{Horvitz:1999:Principles} expanded the literature of design principles that seek to couple direct manipulation of interfaces with automated services by addressing the uncertainty of ascertaining user goals under a cost and benefits framework.
These principles have been explored in increasingly diverse visualization research \cite{Cabrera:2019:FairVis, Endert:2017:StateoftheArt, Liu:2017:BetterAnalysis} since then as designers develop systems that explore the roles they have in augmenting the user experience using VA techniques and systems.
In particular, Ceneda et al. \cite{Ceneda:2017:Guidance} recognize the need for understanding when that role encompasses principles of guidance, and establish a model for designing system initiatives that seek to close a knowledge gap in data analysis.

As mixed-initiative VA research balances human and machine initiatives towards this role of guidance, another role is being investigated: one that concurrently considers the effects of cognitive biases on the analysis process.
Research teams across diverse domains of expertise consider the impacts of cognitive bias regularly as they apply to adoption of VA within their domains \cite{Dimara:2018:Mitigating, Dimara:2016:Attraction, Nussbaumer:2016:Framework, Padilla:2018:DecisionMaking, Wall:2018:Perspectives, Wenskovitch:2020:IinInteraction}.
However, little research has yet been devoted to understanding this interplay of cognitive theory and design of mixed-initiative VA systems as an interdisciplinary framework of future research.
Specifically, such frameworks have the potential to spawn lines of research towards designing and developing tools that can help mitigate cognitive biases and promote fairness -- and lead toward a bias-aware future for mixed-initiative VA systems.
This paper identifies opportunities and challenges that exist towards solving this challenge, and discusses them in context of related work and open questions.

\section{Existing Domain Applications of Visual Analytics Techniques and Systems}
VA systems today see increased usage in a wider variety of contexts than ever before, with researchers being encouraged to expand the scope of contributions to the field into more intellectually diverse areas\cite{Lee:2019:Broadening}. 
The prominence of these systems in the ML community is underscored by their ability to create insight into previously indecipherable streams of information that come from the ``black boxes'' of code powering them\cite{Liu:2017:BetterAnalysis}. 
In particular, the deep learning community is beginning to leverage the power of visually representing relationships and structures embedded in immensely complex neural networks to great success\cite{Hohman:2019:VAinDeepLearning}. 
Urban computing, with its emphasis on large, complex data sets, and urban computing experts have increasingly turned to VA for data exploration, pattern interpretation, and visual learning goals\cite{Zheng:2016:Urban}.
Expanding research avenues that focus on interactive data analysis and exploration are looking toward VA systems to extract human traits encoded in interaction processes \cite{Wenskovitch:2020:IinInteraction}.
These teams investigating machine initiatives in support of analytical reasoning show the impact of adopting VA across ever more diverse domains.
This diversity suggests that there is a great desire for VA systems to explore the boundaries of conventional visual design principles, resist a singular focus on domain-specific development practices, and test the limits of machine-driven data analysis and exploration using VA techniques and systems.

From the perspective of mixed-initiative literature, this desire for smarter autonomous processes and initiatives should be equally mirrored by a focus on understanding the human endeavors that also drive the analysis process.
However, examples in this direction lack the wider adoption seen above, despite bringing much desired insight and contextualization to the domains above.
In other words, the marriage of cognitive theory with VA system design principles remains limited in application, though teams are beginning to show how domains might take advantage of VA in understanding this research area.
Nussbaumer et al. \cite{Nussbaumer:2016:Framework} develop a preliminary framework for addressing specific cognitive biases in the field of criminal analysis aided by VA systems. 
Wall et al. \cite{Wall:2017:Warning} consider the impact of cognitive bias on the analytic process in existing VA systems, and establish a preliminary framework informed by the fields of cognitive theory and decision-making.
Addressing this interdisciplinary approach, Padilla et al. \cite{Padilla:2018:DecisionMaking} suggest a lack of unifying framework limits cross-domain communication of findings informed by cognitive theory, and develop a decision-making model to address this.
While support continues to grow for developing informed autonomous initiatives for VA systems in ML, urban computing, criminal analysis, and other fields, such systems also comprise a role of understanding and addressing the cognitive behaviors of users during the analysis process.

\section{Opportunities to Take Initiative}
We present a collective vision consisting of multiple ideas for identifying opportunities in designing mixed-initiative VA systems across various domains with the goal of mitigating cognitive bias.
Developers of mixed-initiative systems design specific ``initiatives'' that such tools deliver to guide and support users in their tasks.
Our ideas highlight how such initiatives can be grounded in what different communities know about the cognitive processes of users and we discuss them in context of existing VA systems and literature.

\bigskip \noindent \textbf{Cognitive processes of people can be captured and interpreted systematically to identify cognitive traits, tasks, biases, and more.}
Mixed-initiative systems can employ automated initiatives that strive to address the uncertainty in user's goals and attention \cite{Horvitz:1999:Principles}.
Much of the value added by those principles is determined by the uncertain nature of how those initiatives serve the personal goals of the user \cite{Wenskovitch:2020:IinInteraction}.
FairVis \cite{Cabrera:2019:FairVis}, a mixed-initiative VA system for auditing fairness of ML models, addresses such uncertainty outright using domain knowledge provided by the user to inform the automated uncovering of model subgroups relevant to the user.
Other models of interpreting user goals have been developed that can capture interaction provenance and interpret various aspects of user behavior or cognitive traits \cite{Wenskovitch:2020:IinInteraction}.
We believe systems have information about user goals and intent available in the analytic process itself and can thus help guide and support users better if such information is acted on.
When it comes to addressing bias in analysis, such frameworks are critical.
Current methods of bias mitigation rely largely on strategies that reside outside of the systems themselves, such as manual provenance tracking solutions and other structured analytic techniques that help illuminate errors in the cognitive process of analysis \cite{Heuer:2010:Structured}.
Instead, we envision visual analytic systems built to consider a wide design space for how they can respond once biases are detected \cite{Wall:2019:DesignSpace}.
In other words, designing systems that \textit{take initiative} towards correcting these analysis errors when detected.

\bigskip \noindent \textbf{Showing users a history of their process can help promote awareness and reflection of personal bias.}
Inherent in the push for utilizing provenance data is a belief that mixed-initiative systems can act on this information.
Supporting this idea, Ragan et al. \cite{Ragan:2016:CharacterizingProvenance} develop an organizational framework that expands the literature of provenance types and purpose in the analytic process of visualization and data analysis research.
Padilla et al. \cite{Padilla:2018:DecisionMaking} further address the ability of cross-discipline research to inform the user and machine, and contribute an integrative model that acts on decision-making principles informed by cognitive theory used in visualization for comprehension.
The marriage of these critical components of mixed-initiative VA systems provides a framework in which to realize the potential for impacting the cognitive process of users by showing them what they've done already.
Feng et al. \cite{Feng:2017:HindSight} consider this potential in developing HindSight, a visual encoding technique for capturing and presenting user interaction history while analyzing relatively simple visual charts.
Their findings indicate an ability to alter user behavior as a result of promoting awareness of interaction history, a key aspect supporting the potential to integrate cognitive bias mitigation strategies in mixed-initiative VA systems.

\bigskip \noindent \textbf{Systematic interpretation of analytic provenance can illuminate the direction of guidance.}
Looking at user behaviors and the impact of visualizing provenance data during the analytic process strongly suggests support for adopting a design space that can engage with analytic provenance for addressing cognitive bias in user interactions.
Toward this end, we claim that if systems can measure cognitive bias in the analysis process, then they should compute what the user should do in response. 
We are informed by Ceneda et al. \cite{Ceneda:2017:Guidance} that guidance can be characterized and realized as a process for solving problems by knowing where the user is, and where they wish to go, in the analytic process.
Madangopal et al. \cite{Madangopal:2019:AnalyticProvenance} integrate this understanding of guidance with the expanded literature of analytic provenance in visualization above through use-cases in national security, highlighting practical applications for considering provenance-supported methods.
These research avenues reveal an interdisciplinary opportunity to develop bias mitigation strategies using computational techniques that illuminate the direction of guidance during the analysis process.
Wall et al. \cite{Wall:2019:InteractiveBiasMetrics} have captured anchoring bias from user interactions using computational bias metrics integrated directly into a VA system, building towards the goal of developing mixed-initiative approaches for addressing these and other biases.
Computational models which operate over provenance data from the analytic process are beginning to show promise in developing principles and techniques for guiding user behavior, particularly in the context of cognitive bias mitigation.

\section{A Bias-Aware Future}
Research communities are exploring these opportunities for initiative to expand the literature of cognitive theory as it applies to developing mixed-initiative VA systems.
Yet interdisciplinary research is inherently challenging---differing goals, competing funding sources, distinct channels of communication across scientific venues and other considerations can work against the growth of expanding research directions.
However, while challenging, we posit that the next generation of responsive user interfaces will come from such cross-domain approaches and work towards a bias-aware future for mixed-initiative VA systems. 
We want to understand how such systems can push, coax, and guide users towards a bias-aware workflow between machine and human, balancing embedded bias mitigation strategies with increasing productivity for users. 
Towards this goal, there are several open challenges and questions to address.

\bigskip \noindent \textbf{How can bias-aware design reconcile mixed-initiative approaches with diverse goals?}
Many current VA systems employ fixed heuristics or models to determine which actions or initiatives should be taken when certain user actions are recognized.
For example, they may guide users towards improving the accuracy of prediction models, clustering results, or other well-defined goals in ML applications.
Consider, however, the implications of systems designed to also improve the analysis process by mitigating potential bias, or otherwise guiding users towards goals other than task performance or accuracy.
We understand that perspectives on human bias in VA \cite{Wall:2018:Perspectives} reveal a number of ways that bias manifests in practice, intentionally or otherwise, that certainly impact these design considerations.
Yet, at the end of the day, we still aim to design VA systems in service of user goals without inhibiting performance.

This juxtaposition of diverse goals inspires challenges to overcome and questions to be answered:
\begin{itemize}[topsep=1mm, itemsep=0mm, parsep=1mm, leftmargin=4mm]
    \item Are model accuracy and bias mitigation always in opposition, or can they instead support one another? Perhaps traditional performance metrics need to be revisited to include aspects of fairness or bias?
    \item How should the treatment of such trade-offs between accuracy and interpretation influence the implementation of bias-aware principles on either end?
    \item How can the growing desire for faster computation be reconciled with additional layers of bias recognition and mitigation? Faster and more accurate may not be better than bias-aware or fair.
    \item At the end of the day, how can design find the balance along this spectrum between productivity and proclivity for bias?
\end{itemize}
These and other questions involving multiple fields of research will continue to pervade the conversation surrounding the future of mixed-initiative VA research.
As such, we hope this need for balance can be discussed across disciplines and addressed with informed opportunities to take initiative.

\bigskip \noindent \textbf{How far should systems go to balance multiple goals and initiatives?}
When designing systems with numerous diverse goals or initiatives, situations will arise that require deciding between potentially competing goals.
The application of mixed-initiative principles introduces several patterns of design that sacrifice opportunities to engage with cognition in the analytic process in service of more immediate user goals \cite{Wall:2019:DesignSpace}.
To what degree then should mixed-initiative VA systems strive to balance the influences of cognitive biases with developed mixed-initiative user interface principles?

Consider once again the prior example of interfaces for predictive modeling, clustering, or other well-defined goals:
\begin{itemize}[topsep=1mm, itemsep=0mm, parsep=1mm, leftmargin=4mm]
    \item Clustering metrics exist which can inform how systems should guide users towards more accurate clusters; however, what if such systems also had metrics that guide users towards reducing the potential for biased interactions?
    \item Further, what if the guidance or initiative these systems recommended to users differed between these two example goals?
    \item Can existing frameworks account for the integration of metrics informed by cognitive theory with existing interaction tracing techniques?
    \item To what extent would these cognitive bias influences in mixed-initiative applications correlate with interaction patterns and traces consistent with existing VA literature?
    \item How would such testing be implemented in practice, and to what extent be made available to the user, if at all?
\end{itemize}
We believe these limitations in cross-domain communication can be a motivation to designers in VA, ML, and other research areas, helping them consider how informed bias-aware initiatives in mixed-initiative VA systems might take shape in light of interdisciplinary research goals.

\section{Conclusion}
If mixed-initiative VA systems are going to continue to support the user, then opportunities to take initiative should provide a re-contextualized way forward from our evolving understanding of cognitive bias in the analytic process.
This paper identifies a number of open research directions for visual analytics, where bias-awareness and fairness play a central role. 
It is our hope that the discussion of these opportunities to create a bias-aware future invigorate the community to spawn new lines of research towards addressing these important challenges.

\acknowledgments{This work was supported in part by the National Science Foundation grant IIS-1813281.}

\bibliographystyle{abbrv-doi}

\bibliography{references}

\end{document}